# IMPROVING THE PRESENTATION AND UNDERSTANDING OF RISK MODELS

| | |
|---|---|
| Authors: | Ralph H. Stern PhD, MD |
| | Dean E. Smith PhD |
| | Hitinder S. Gurm MD |
| e-mail: | stern@umich.edu |
| Address: | Cardiovascular Medicine |
| | University of Michigan |
| | 24 Frank Lloyd Wright Drive, Lobby A |
| | Ann Arbor, MI 48105 |
| Pages: | 14 |
| Figures: | 7 |

ABSTRACT

The key concepts (calibration, discrimination, and discordance) important in understanding and comparing risk models are best conveyed graphically.  To illustrate this, models predicting death and acute kidney injury in a large cohort of PCI patients differing in the number of predictors included are presented.  Calibration plots, often presented in the current literature, present the agreement between predicted and observed risk for deciles of risk.  Risk distribution curves present the frequency of different levels of risk.   Scatterplots of the risks assigned to individuals by different models show the discordance of the individual risk estimates.  Increasing the number of predictors in these models produce increasingly disperse and progressively skewed risk distribution curves.  These resemble the lognormal distributions expected when risk predictors interact multiplicatively.  These changes in the risk distribution curves correlate with improved measures of discrimination.

INTRODUCTION

Clinicians are frequently challenged to interpret the literature on risk models and risk factors. A major reason is that publications on predictive models have largely focused on numerical presentations of their development and/or comparison. And the experts may disagree on what numerical measures should be used, what they mean, and their clinical significance.[1]

Calibration curves, scatterplots that compare predicted and observed risk in each decile of risk, are often presented. Recently the value of graphical presentations of risk distribution curves or predictiveness curves has been emphasized.[2,3] The former is a graph of the frequency of risk versus risk and the latter a graph of risk versus the cumulative frequency of risk. Pepe, Gu, and Morris have written that "Displaying risk distributions is a fundamental step in evaluating the performance of a risk prediction model, a step that is often overlooked in practice."[4] Another graphical presentation, a scatterplot presenting the discordance of individual risk estimates when two models are compared, was introduced by Lemeshow et al.[5] and has recently been rediscovered by Pencina et al.[6]

In this paper we use these graphical presentations to characterize models differing in the number of risk factors providing insights not available from numerical presentations.

METHODS

The study cohort for our analysis included patients undergoing Percutaneous Coronary Intervention (PCI) in years 2007 and 2008 in a large regional registry of contemporary PCI. The details of the Blue Cross Blue Shield of Michigan Cardiovascular Consortium registry (BMC02) and of the data collection process have been described elsewhere.[7, 8, 9, 10] Briefly procedural data on all patients undergoing elective and non-elective PCI at the 31 participating hospitals is collected using standardized data collection forms. Baseline data include clinical, demographic, procedural, and angiographic characteristics as well as medications used before, during, and after the procedure, and in-hospital outcome. All data elements have been prospectively defined and the local institutional review board at each institution approved the protocol. The data is collected by a dedicated staff member and forwarded to the coordinating center. Medical records of all patients undergoing coronary artery bypass grafting (CABG), and of patients who died in the hospital are reviewed by auditors from the coordinating center to ensure data accuracy. A further 2% of cases are randomly selected for audit.

Two outcomes were considered: all-cause in-hospital death (death) and acute kidney injury (AKI). AKI is an in-hospital outcome defined as "peak minus baseline creatinine ≥0.5 mg/dL", with peak creatinine measured in-hospital before discharge. Death was less common (1.10%) than AKI (3.40%).

Of the total cohort, 60,654 patients were included in the death model while 47,775 patients were included in the AKI model. Of the 13,209 patients who were not included in the AKI model, 1238 were on dialysis prior to the procedure while 10595 patients were excluded due to absence of serum creatinine before or after the procedure and 1756 were excluded due to absence of body weight.

We developed a series of models for death and AKI strictly for illustrative purposes. Our goal was to show the effect of increasing numbers of predictors on the risk distribution curve and associated measures. Full models (model 5) for death and AKI were developed by ascending stepwise logistic regression. Then a series of nested submodels (models 1, 2, 3, and 4) with increasing numbers of predictors were constructed for death and AKI. Model 1 had only had age and gender. Subsequently additional predictors were added to generate models 2, 3, and 4, with the strongest predictors added first. Predictors were categorical variables, with the exception of baseline creatinine in the AKI models. The predictors in the death model were limited to those available prior to PCI, while the predictors in the AKI model also included procedural and angiographic data.

Statistical support for this project was constrained by time, so additional analyses and/or alternative graphics could not be generated.

## RESULTS

Tables 1 and 2 summarize the 5 models for death and AKI, respectively. As more predictors are included in the models, measures of goodness of fit and discrimination are progressively improved. Cardiogenic shock was the strongest predictor for death and AKI.

For death, although model 1 has a significant Hosmer-Lemshow statistic, the other models are calibrated by this metric. Figure 1 shows the calibration plots lie near the line of identify for the 5 death models, which supports the calibration of the models. For AKI, however, all but model 1 have significant Hosmer-Lemeshow statistics. Again, the calibration plots in figure 2 support the calibration of the models. Significant Hosmer-Lemeshow tests with small departures from a proposed model with large data sets is well recognized.[11]

Figures 3 and 4 present the risk distribution curves for the 5 models for death and AKI, respectively. Even model 1 with only age and gender produces distributions with substantial dispersion around the mean risks. (for death: 668 in 60,656 subjects=0.011013, for AKI 1615 in 47,446 subjects without missing data=0.034039)  As additional predictors are added, the curves become increasingly disperse. In addition the curves become increasingly skewed.

More disperse risk distribution curves assign fewer cases to the $1^{st}$ and more cases to the $10^{th}$ decile, reflecting their improved discrimination.   Of the 668 deaths, the number of cases in the $1^{st}$ decile/$10^{th}$ decile were 35/183, 6/472, 3/460, 2/506, and 3/510 in models 1, 2, 3, 4, and 5, respectively.  Of the 1615 cases of AKI, the number of cases in the $1^{st}$ decile/$10^{th}$ decile were 64/290, 27/752, 24/780, 23/838, and 22/849 in models 1, 2, 3, 4, and 5, respectively.

Numerical tables, calibration plots, and risk distribution curves describe how addition of predictors improves the risk stratification of the population, but not their effects on individuals.   This is best appreciated by inspecting a scatterplots of the individual risk estimates from two models.  Figures 5 and 6 do this for models 4 and 5 for death and AKI, respectively.  These two models were the closest in terms of risk stratification of the population, but clearly give different results at the individual level.

In theory, risk distribution curves generated from multiple risk factors that interact multiplicatively should be lognormal.[12]  In figure 7 simulated lognormal risk distribution curves with mean risks of 0.011 and 0.034, the mean risks of death and AKI, respectively, but differing dispersion are presented for comparison to the risk distribution curves of figures 3 and 4.  Curves with low dispersion are symmetrical and centered on the mean.  Curves with higher dispersion are progressively skewed to the right.

DISCUSSION

We agree with Jaynes, who wrote "The First Commandment of scientific data analysis publication ought to be: 'Thou shalt reveal thy full original data, unmutilated by any processing whatsoever.'"[13] Before calculating statistics, especially on categorized data, readers should be given an opportunity to visualize the data.

Numerical results do not provide the insight provided by risk distribution curves. The latter present the location, dispersion, and shape of the risk distribution curve. Comparison of risk distributions differing in the number of predictors allows a better understanding of the modeling process.

Nomenclature is confusing as both calibration and discrimination are referred to as accuracy. However they are completely different conceptually. Calibration refers to the agreement between predictions and observations. But even a model with no predictors can have perfect calibration. Such a model would assign everyone the mean risk and, if the observed risk in the population matched this prediction, then the model would be perfectly calibrated. Addition of risk factors to a model should not alter the agreement between observations and predictions. Thus accuracy, defined as calibration, is not a function of the number of predictors in a model. Calibration plots presenting observed versus predicted risk for each decile of risk allow for a graphical evaluation of the agreement.

Although additional risk factors do not provide improved calibration or accuracy, they do provide improved discrimination. This is measured by the c-statistic or area under the ROC curve, which are measures of the overlap between the risk distribution curves for cases and controls, both of which are fully determined by the population risk distribution. Improved discrimination reflects the dispersion of the risk distribution curve. Narrow risk distribution curves assign patients who will have events and those who won't have events similar probabilities of an outcome. In this instance there must be substantial overlap of the derived risk distribution curves for cases and controls. On the other hand, as additional risk factors are included, broader risk distribution curves result in cases and controls being assigned increasingly different probabilities of an outcome. This is readily appreciated when a composite plot of risk distribution curves generated by models differing in the number of predictors is presented.

Presenting risk distribution curves separately for cases and controls has been advocated.[14] We agree this well depicts the discrimination of the two groups. However we favor the population risk distribution curve as, in theory, the risk distribution curves for cases and controls can be calculated directly from the risk distribution curve for the population.[15]

An additional feature apparent from the composite plot is that the distributions also become increasingly asymmetric as additional predictors are included. When a sufficient number of risk factors interact multiplicatively, the expected distribution of risk in the population is lognormal.[12] We suggest that that is the case with this series of models. Narrow lognormal curves are symmetric and resemble normal distributions. But increasing dispersion is associated with increasingly asymmetric lognormal curves. That simulated lognormal curves with the same mean risks as observed for death and AKI appear similar supports the suggestion of lognormality.

But even the addition of risk distribution curves to numerical presentations provides no insight into how individuals are characterized by different models. This is best appreciated from a scatterplot of the calculated risks for individuals derived from two models, an approach introduced by Lemeshow et al. When models are nested, as is the case here, this discordance may be less than when there are fewer shared predictors. Nonetheless, even when addition of predictors has little or no impact on the dispersion of the risk distribution curve, there can be significant disagreement. When the risk distribution is split into categories, disagreement has been termed reclassification. However when the models are calibrated and their risk distributions are similar, this reclassification has little clinical significance.

The presentation and understanding of risk models would be improved by including risk distribution curves in all publications. Graphical presentation of the data prior to categorization and/or statistical analysis is standard practice in other areas of science and medicine. Risk distribution curves directly depict the dispersion of the risks assigned by a model. That measures of discrimination reflect this dispersion is an important point that has been overlooked because of the lack of graphical presentation. When more than one model is utilized, comparison of risk distribution curves allow the observer to assess whether they differ in location (demonstrating a difference in calibration) or dispersion (demonstrating a difference in discrimination). Since risk distribution curves are not presented, there has been no attention paid to their shape and the influence of increasing the number of included predictors on their shape. A normal distribution cannot be assumed and the skewing of distributions would be important information for modeling the clinical benefit of adding additional predictors to a model

Scatterplots of the predicted risk for individuals from two models, originally proposed by Lemeshow et al.[5] and more recently suggested by Pencina et al.,[6] should also be presented in publications. As for risk distribution curves, this should precede categorization and/or statistical analysis.

The clinical goal of model development is to identify risk factors that produce optimal population risk stratification, which means identification of subpopulations that are as large as possible and differ as much as possible from each other and that are calibrated. Intuitively, this should correspond to a risk distribution curve as disperse as possible.

# Table 1

## DEATHMODELS

| | MODEL 1 | | MODEL 2 | | MODEL 3 | | MODEL 4 | | MODEL 5 | |
|---|---|---|---|---|---|---|---|---|---|---|
| C statistic | 0.662 | | 0.883 | | 0.901 | | 0.912 | | 0.921 | |
| H-L statistic | 0.0004 | | 0.1976 | | 0.0981 | | 0.1771 | | 0.1800 | |
| AIC | 7128 | | 5296 | | 5139 | | 4943 | | 4861 | |
| | EST | OR | EST | OR | EST | OR | EST | OR | EST | OR |
| Intercept | -7.8939 | | -9.9657 | | -10.2236 | | -10.3139 | | -10.1431 | |
| Gender | 0.1179 | 1.125 | 0.1010 | 1.106 | 0.1934 | 1.213 | 0.2964 | 1.345 | 0.2579 | 1.294 |
| Age | 0.0494 | 1.051 | 0.0580 | 1.060 | 0.0545 | 1.056 | 0.0526 | 1.054 | 0.0442 | 1.045 |
| Shock/MI | | | 2.5382 | 12.656 | 2.3521 | 10.507 | 1.9117 | 6.764 | 1.8664 | 6.465 |
| MI | | | 1.8468 | 6.339 | 1.5433 | 4.680 | 0.9864 | 2.682 | 0.9682 | 2.633 |
| Arrest | | | 1.6206 | 5.056 | 1.5201 | 4.573 | 1.4194 | 4.135 | 1.4342 | 4.196 |
| LVEF<50 | | | | | 1.2130 | 3.364 | 1.0734 | 2.925 | 0.9693 | 2.636 |
| Crt>2 | | | | | | | 1.270 | 3.561 | 0.9486 | 2.582 |
| Crt 1.5-2 | | | | | | | 0.7717 | 2.163 | 0.6504 | 1.916 |
| Emergency PCI | | | | | | | 1.1210 | 3.068 | 1.3090 | 3.702 |
| Valve disease | | | | | | | | | 0.6580 | 1.931 |
| Anemia | | | | | | | | | 0.5643 | 1.758 |
| PVD/CVA | | | | | | | | | 0.3599 | 1.433 |
| Hx CHF | | | | | | | | | 0.1923 | 1.212 |

**Table 2**

**AKI MODELS**

| | MODEL 1 | | MODEL 2 | | MODEL 3 | | MODEL 4 | | MODEL 5 | |
|---|---|---|---|---|---|---|---|---|---|---|
| C-statistic | 0.645 | | 0.797 | | 0.812 | | 0.832 | | 0.836 | |
| HL statistic | 0.8284 | | 0.0016 | | 0.0209 | | 0.001 | | 0.0154 | |
| AIC | 13680 | | 12046 | | 11806 | | 11458 | | 11361 | |
| | EST | OR | EST | OR | EST | OR | EST | OR | EST | OR |
| Intercept | -6.1013 | | -7.8511 | | -7.079 | | -7.8573 | | -7.5164 | |
| Gender | 0.3207 | 1.378 | 0.37370 | 1.452 | 0.3553 | 1.427 | 0.3689 | 1.446 | 0.2493 | 1.283 |
| Age | 0.0391 | 1.040 | 0.0385 | 1.039 | 0.0332 | 1.034 | 0.0278 | 1.028 | 0.0181 | 1.018 |
| Shock/MI | | | 1.8004 | 6.052 | 1.5306 | 4.621 | 1.4151 | 4.117 | 1.2439 | 3.469 |
| MI | | | 1.1698 | 3.221 | 0.8854 | 2.424 | 0.7995 | 2.224 | 0.7226 | 2.060 |
| DM | | | 0.8698 | 2.386 | 0.7935 | 2.211 | 0.6121 | 1.844 | 0.6193 | 1.858 |
| Pre Crt | | | 0.6147 | 1.849 | 0.5340 | 1.706 | 0.4656 | 1.593 | 0.3205 | 1.378 |
| Anemia | | | | | 0.7977 | 2.220 | 0.6478 | 1.911 | 0.614 | 1.849 |
| Emergency PCI | | | | | 0.6585 | 1.932 | 0.7243 | 2.063 | 0.6020 | 1.826 |
| Valve disease | | | | | | | 0.5455 | 1.725 | 0.5280 | 1.696 |
| LVEF<50 | | | | | | | 0.5356 | 1.708 | 0.5285 | 1.696 |
| Hx CHF | | | | | | | 0.5294 | 1.698 | 0.5041 | 1.655 |
| PVD/CVA | | | | | | | 0.4384 | 1.550 | 0.4198 | 1.522 |
| Obese | | | | | | | 0.2962 | 1.345 | 0.4694 | 1.599 |
| Overweight | | | | | | | 0.1319 | 1.141 | 0.2295 | 1.258 |
| Crt Cl 1-30 | | | | | | | | | 0.8138 | 2.2257 |
| Crt Cl 30-59 | | | | | | | | | 0.5240 | 1.689 |
| Crt Cl 60-89 | | | | | | | | | 0.3399 | 1.405 |
| Cardiac arrest | | | | | | | | | 0.4089 | 1.505 |
| Stenosis 70 | | | | | | | | | 0.3880 | 1.474 |
| COPD | | | | | | | | | 0.2790 | 1.322 |
| Thrombus | | | | | | | | | 0.2558 | 1.291 |
| Vess dis 70 | | | | | | | | | 0.2050 | 1.227 |
| H HYP | | | | | | | | | 0.1735 | 1.190 |
| Calcification | | | | | | | | | 0.1409 | 1.151 |
| Hx PTCA | | | | | | | | | -0.1761 | 0.839 |
| CABG | | | | | | | | | -0.3970 | 0.672 |

**FIGURES**

Figure 1. Death Calibration Plots.

    Model 1, orange; model 2, blue; model 3, green; model 4, red; and model 5, black

Figure 2. AKI Calibration Plots

    Model 1, orange; model 2, blue; model 3, green; model 4, red; and model 5, black

Figure 3. Death Risk Distribution Curves

    Model 1, orange; model 2, blue; model 3, green; model 4, red; and model 5, black

Figure 4. AKI Risk Distribution Curves

    Model 1, orange; model 2, blue; model 3, green; model 4, red; and model 5, black

Figure 5. Death Discordance Plot for Model 4 versus Model 5

Figure 6. AKI Discordance Plot for Model 4 versus Model 5

Figure 7. Simulated Lognormal Risk Distribution Curves

**FIGURE 1**

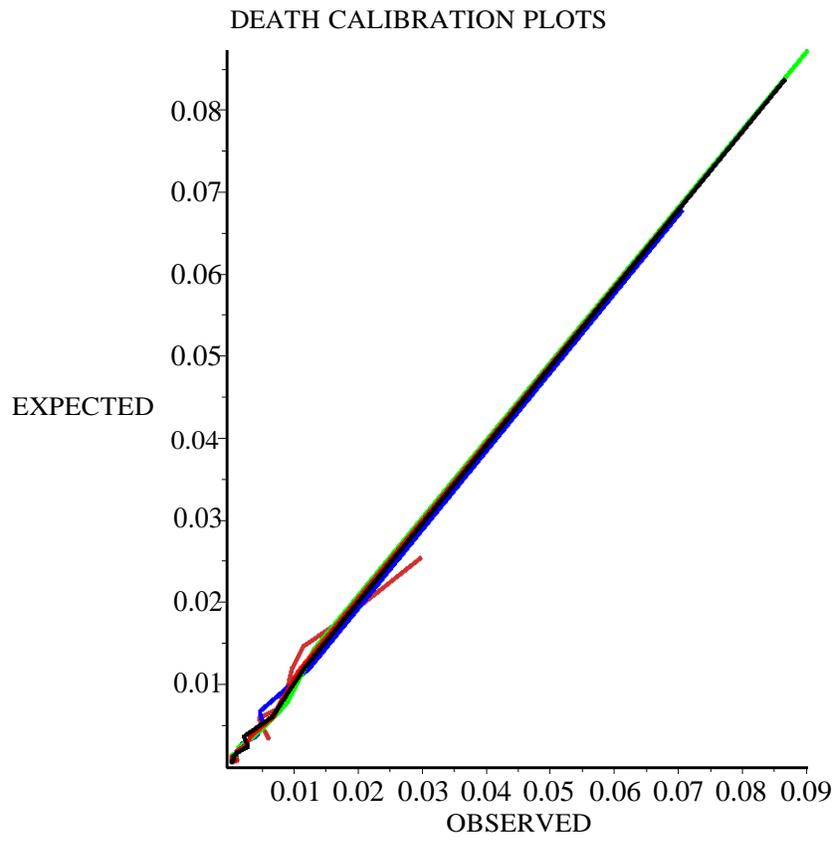

**FIGURE 2**

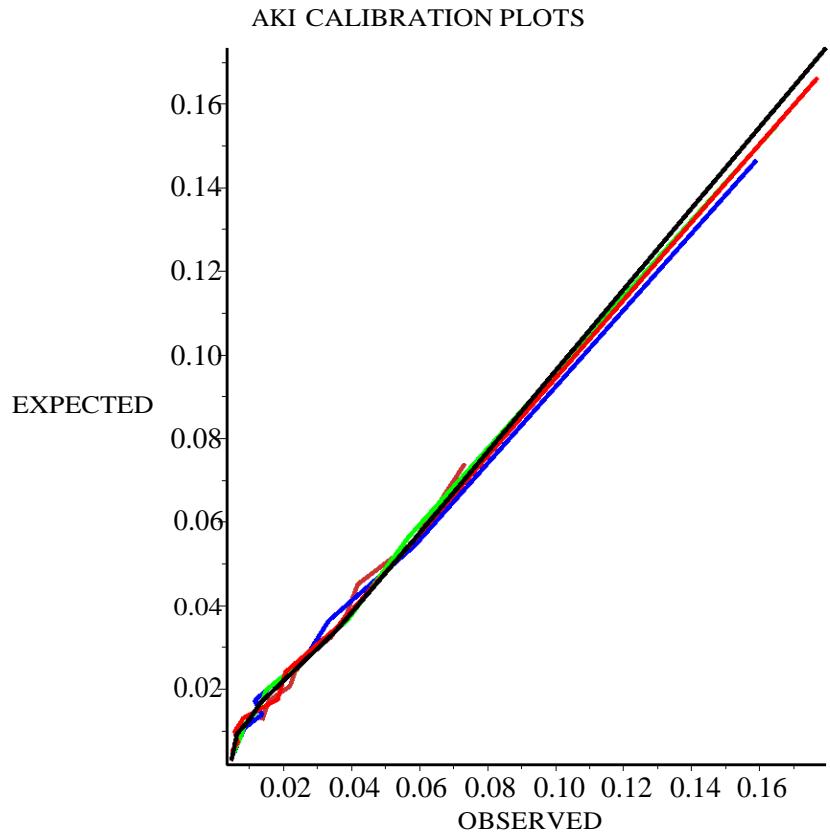

**FIGURE 3**

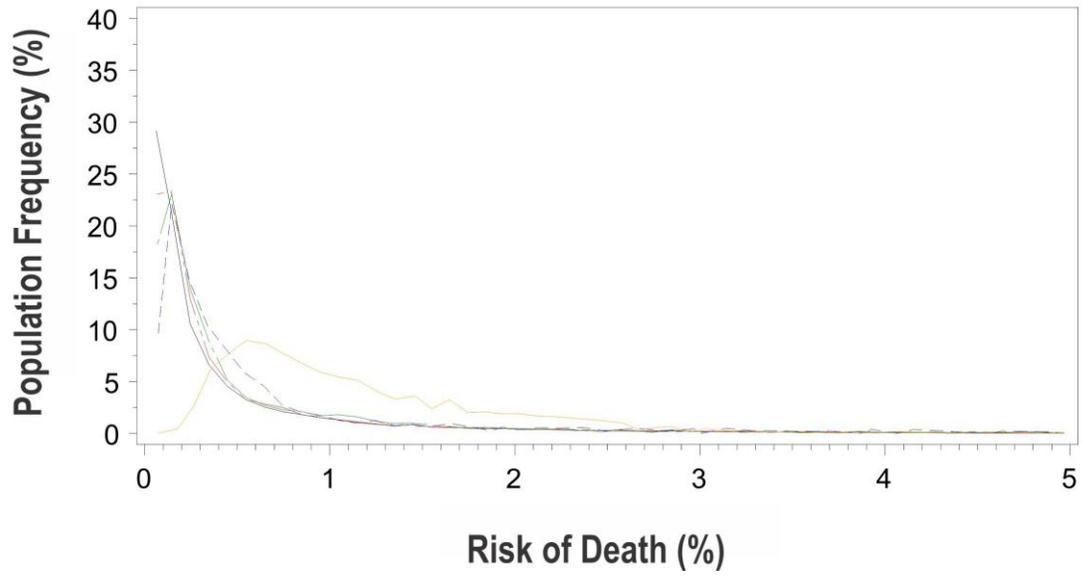

**FIGURE 4**

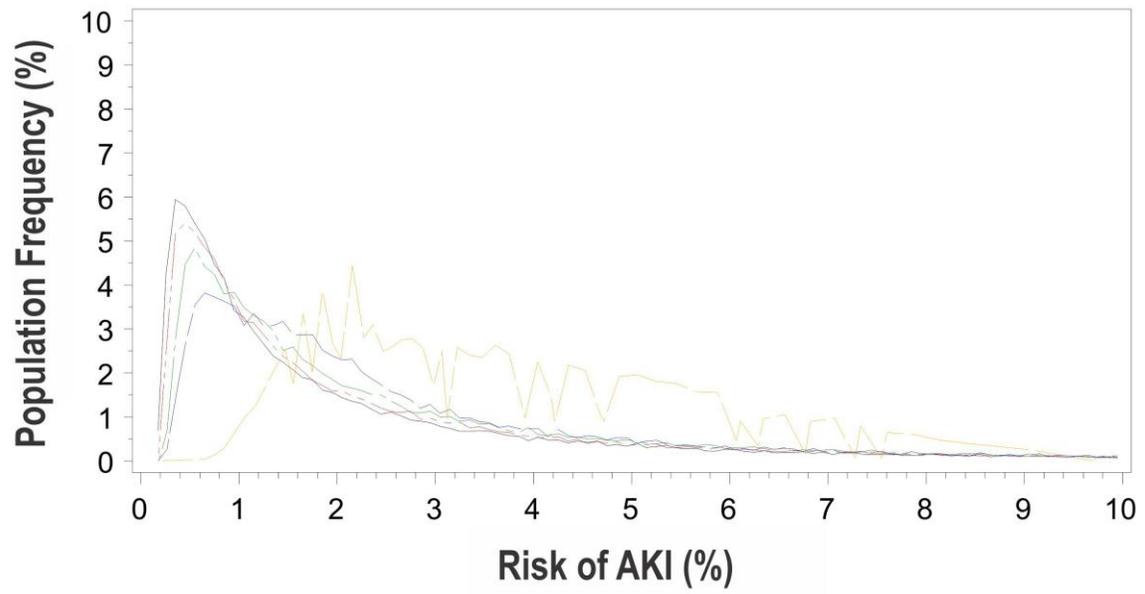

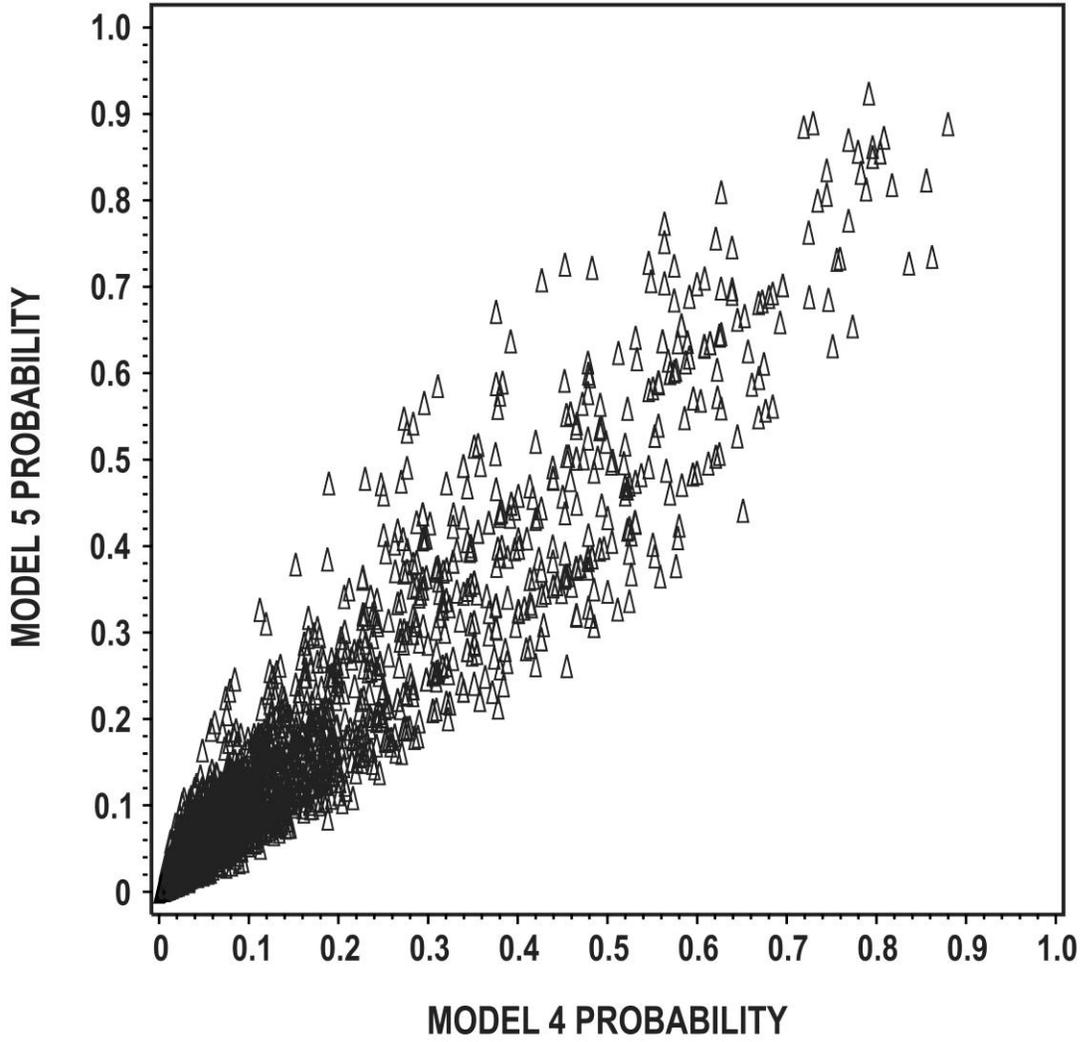

**FIGURE 6**

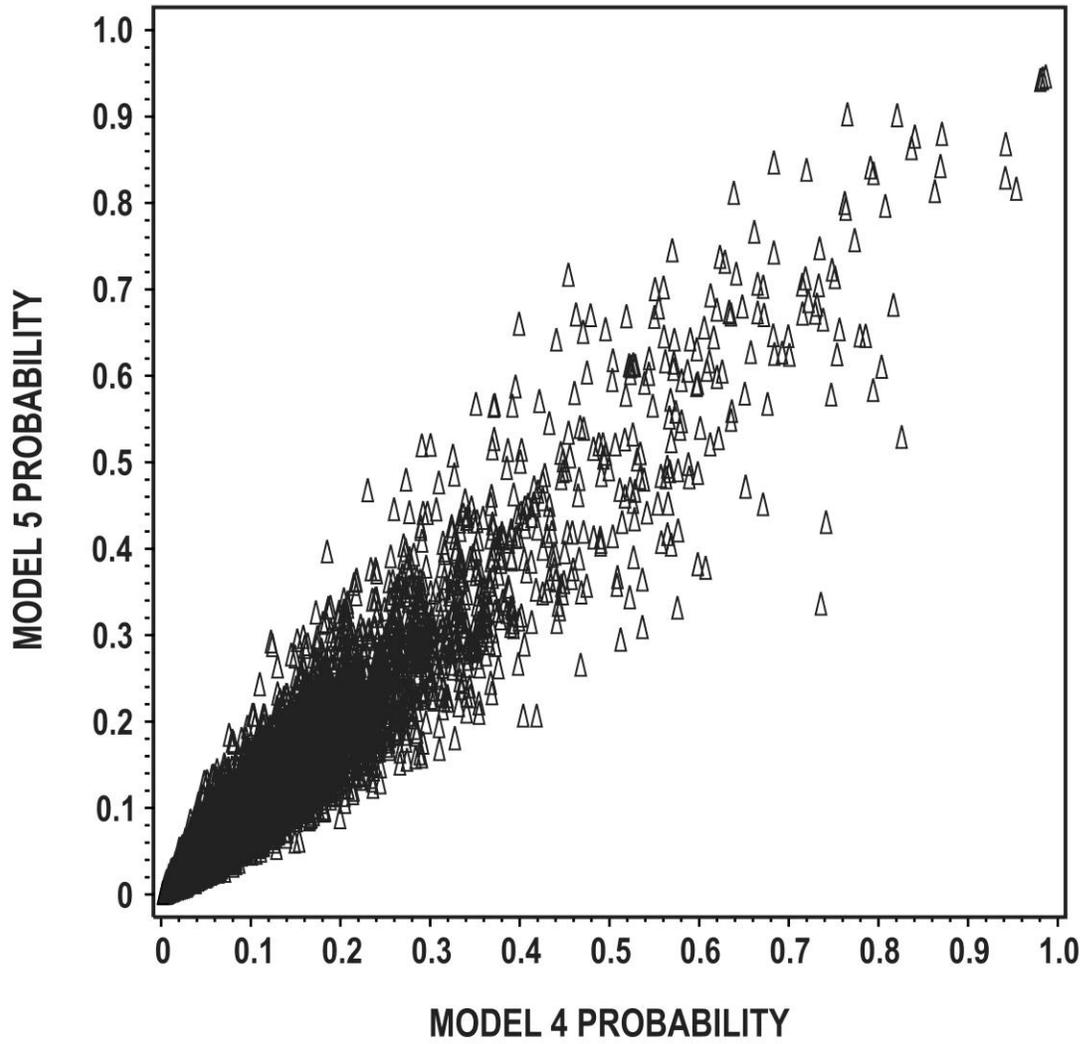

**FIGURE 7**

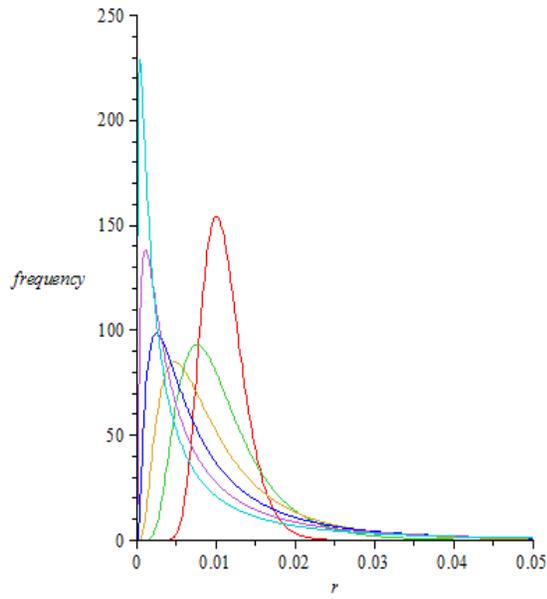

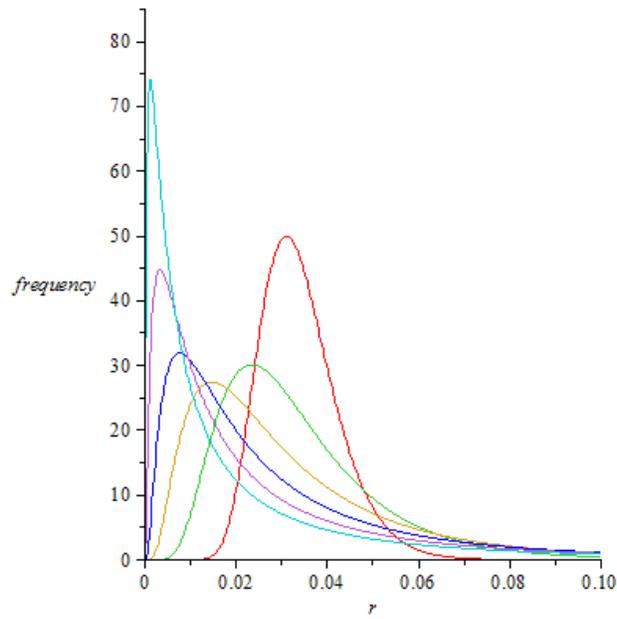

Simulated Lognormal Risk Distribution Curves with Mean Risks of 0.011 (above) and 0.034 (below)